# Analytic Trajectories for Mobility Edges in the Anderson Model

Wolfram T. Arnold and Roger Haydock

Department of Physics and Materials Science Institute

University of Oregon, Eugene OR 97403-1274, USA

## Abstract

A basis of Bloch waves, distorted locally by the random potential, is introduced for electrons in the Anderson model. Matrix elements of the Hamiltonian between these distorted waves are averages over infinite numbers of independent site-energies, and so take definite values rather than distributions of values. The transformed Hamiltonian is ordered, and may be interpreted as an itinerant electron interacting with a spin on each site. In this new basis, the distinction between extended and localized states is clear, and edges of the bands of extended states, the mobility edges, are calculated as a function of disorder. In two dimensions these edges have been found in both analytic and numerical applications of tridiagonalization, but they have not been found in analytical approaches based on perturbation theory, or the single-parameter scaling hypothesis; nor have they been detected in numerical apporaches based on scaling or critical distributions of level spacing. In both two and three dimensions the mobility edges in this work are found to separate with increasing disorder for all disorders, in contrast with the results of calculations using numerical scaling for three dimensions. The analytic trajectories are compared with recent results of numerical tridiagonalization on samples of over $10^9$ sites. This representation of the Anderson model as an ordered interacting system implies that in addition to transitions at mobility edges, the Anderson model contains weaker transitions characterized by critical disorders where the band of extended states decouples from individual sites; and that singularities in the distribution of site energies, rather than its second moment, determine localization properties of the Anderson model.

1.  Disorder and the Anderson Model

Why electrons move freely in some solids, but not others, is a fundamental problem of solid state physics.  Bloch's theorem says that for periodic potentials electronic states spread evenly over the entire system, but Bloch's theorem does not say how far from periodic the potential must be for electronic states to be localized in finite regions of the system.  This is the question addressed by the Anderson model[1] for disorder and this paper.

While disorder comes in many forms which are difficult to characterize, the disorder captured in the Anderson model is simple but sufficient to localize electrons.  In this model, the wave function for each electron is a linear combination of orthonormal orbitals $\{\phi_\alpha\}$ centered on sites $\{\mathbf{R}_\alpha\}$ of a $d$-dimensional periodic lattice, taken here to be linear, square, or simple cubic for one, two, or three dimensions respectively.  The Anderson Hamiltonian has the same matrix-element, h, for hopping between nearest-neighbor orbitals with all other hopping matrix-elements zero.  The disorder is in the energies $\{\epsilon_\alpha\}$ of the orbitals which are taken independently for each site from the same probability distribution $\rho(\epsilon)$, often a top hat distribution of width W.  In terms of operators $\{c_\alpha\}$ which annihilate electrons in the orbitals $\{\phi_\alpha\}$, the Anderson Hamiltonian is,

$$H = \sum_\alpha [\epsilon_\alpha c_\alpha{}^\dagger c_\alpha + h \sum_\delta c_\alpha{}^\dagger c_{\alpha+\delta}], \tag{1}$$

where site $\alpha+\delta$ is a nearest neighbor of site $\alpha$.

When $\rho(\epsilon)$ is a delta-distribution, the $\{\epsilon_\alpha\}$ all have the same value, so the Hamiltonian in Eq. 1 is periodic, and the eigenstates of H are extended by Bloch's theorem.  If $\rho(\epsilon)$ has breadth, the $\{\epsilon_\alpha\}$ take different random values and Bloch's theorem no longer applies to H.  Experience shows that the most important property of $\rho(\epsilon)$ is its width W which has been characterized in various ways, but for this work is taken to be *the difference between the extremal values of* $\epsilon$.  As W increases from zero, H develops eigenstates which are less like Bloch states and more like molecular combinations of orbitals or even single orbitals.  The extremal eigenvalues of H are at the Lifshitz edges[2], $\pm(2d\text{h}+\text{W}/2)$, where $d$ is the dimension of the lattice (linear, square, or cubic). The simplest distinction between different kinds of states is that the weight of extended states is distributed over an infinite number of orbitals, in contrast to localized states which have negligible weight outside a finite number of orbitals. Other criteria such as normalization or time-reversal properties can also be used to distinguish different kinds of states. When W is small but non-zero and $d$ is large, states with qualitatively different localization properties form bands whose extremal energies are called mobility edges [2]. In addition to the mobility edges, there are Anderson transitions[1] at which the electrons cease to escape from individual sites for critical values of W approximately equal to the width of the band with no disorder.

The main problem addressed in this paper is how the energies of mobility edges vary with disorder W, their trajectories, for lattices of different dimensionality $d$. Of particular interest is what happens to these trajectories near Anderson transitions. When W is zero, the states of H are all extended forming a band with extremal energies at $\pm 2d$h. Single defects can produce localized states outside the band of extended state, so in this sense, $\pm 2d$h are also mobility edges for W zero. As W increases, the Lifshitz edges move apart linearly with W irrespective of dimensionality, while the behavior of mobility edges seems to depend on dimensionality. In three dimensions for W small compared to h, it is generally agreed that there are mobility edges separating exponentially localized states near the band edges from extended states near the band center, and that these edges move apart sub-linearly in W. For stronger disorder, numerical scaling calculations [3,4,5,6] support the picture in which the trajectories of the mobility edges curve inward and meet at the center of the band at the critical disorder for the Anderson transition. In two dimensions for W small, analytic and numerical applications of the recursion method[7,8,9] find evidence of mobility edges separating exponentially localized states at the band edges from power-law localized states near the band center, and evidence that these edges also move apart sub-linearly in W. Other approaches [10,11,12] find only exponentially localized states in two-dimensions for non-zero W, and no evidence of mobility edges. In one dimension, it is generally agreed that the states coupled to site-orbitals are all exponentially localized [2] and that the localization lengths of these states vary smoothly with energy across the entire band.

What is new in the work reported below is the use of extended states rather than localized orbitals as a basis for the Anderson Hamiltonian.  One might think that this is like using plane waves instead of tight-binding orbitals as the basis for a band structure calculation, and that the differences in results would be small and easy to understand.  However the advantage of an extended basis is that an infinite number of independent site-energies contribute to each matrix-element between extended states and so the disorder averages out of these matrix-elements.  Furthermore, because the Anderson Hamiltonian is infinite but not periodic, it can have many kinds of eigenstates with very different expansions in localized and extended bases.  The most surprising consequence of this new approach, which is explained in Secs. 2 and 3, is that for even very large disorder there is always a band of extended states in which individual site-orbitals have a weight which depends on the dimensionality and disorder.  When $d$ is large and W small, this weight is non-zero, but for large W or small $d$, the weight goes to zero and the extended states become decoupled from the site-orbitals which participate in a band of localized states degenerate with the extended states.  These extended states are not a subtle effect, but are straightforward linear combinations of the distorted waves constructed below.  The extremal energies of these bands of extended states coincide with the mobility edges seen in previous work [7,8], as is discussed in Sec. 4.

The existence of extended states for all disorders seems at first to conflict with the existence of the Anderson transition which arises both in the recursion method [13] and other methods[14].  The conflict is resolved in Sec. 5 by noting that for sufficiently large disorder, the site-orbitals become decoupled from the extended states in the sense that the weight of a single orbital, integrated over the entire band of extended states, goes to zero.  This contradicts the picture of the Anderson transition as taking place when the mobility edges meet at zero energy[2,11], but not with the existence of Anderson transitions.

While most of this paper deals with distributions $\rho(\varepsilon)$ having finite width W, at the end of Sec. 5 there is a brief discussion of the Anderson model with unbounded disorder, for example a Gaussian distribution, arguing that its infinite width leads to an Anderson transition at zero disorder.

2.  A Basis of Distorted Waves

Extended states can only be distinguished from localized states in infinite systems because finite combinations of orbitals are always localized.  In an infinite system the localization properties of a state are determined by the asymptotic behavior of its components, however a continuum of qualitatively different asymptotic behaviors is possible.  In disordered systems, this is further complicated by random fluctuations in the states, due to the random potential.  The main impediment to understanding the localization properties of systems such as the Anderson model is the difficulty of determining the asymptotic properties of states.

One consequence of their different asymptotic behavior is that localized and extended states span the site-orbitals in different ways.  A finite number of orbitals contribute significant relative weight to each localized state, while no individual site-orbitals contribute significant relative weight to an extended state because the extended states are distributed over an infinite number of orbitals.  As a result of the qualitative difference in the way orbitals contribute to localized and extended states, it is necessary to construct these states differently.

Most studies of the Anderson model concentrate on the properties of the localized states, using the site-orbitals of the model as a basis.  The approach adopted here is to concentrate on the extended states by expanding them in an extended basis, a basis made of infinite linear combinations of site-orbitals.  However, to make the problem tractable in the extended representation, the Anderson Hamiltonian must couple each element of the extended basis to only a few other elements, a sparse representation.  To achieve this, the extended basis must consist of waves which are distorted locally to adapt to the disordered potential.

In addition to emphasizing the extended states, the distorted-wave basis has the advantage of averaging out the disorder in the matrix-elements of the Hamiltonian.  Each basis element is a linear combination of an infinite number of site-orbitals, so the matrix-elements between basis elements are averages of an infinite number of site-energies.  Provided that the distributions of site-energies satisfy the requirements of the central limit theorem, the second moments of the distributions of matrix-elements are zero, and so each matrix-element takes a definite value rather than a distribution of values.

Construction of this basis of distorted waves begins with a single, undistorted Bloch wave, for which it is simplest to take the constant wave,

$$\Phi_{\mathbf{0}} = \sum_{\alpha} \phi_{\alpha}, \tag{2}$$

where the waves are indexed by the degree of their dependence on the energies of the different sites in the lattice.  For $\Phi_{\mathbf{0}}$, the components of the index $\mathbf{0}$=(0,0,0,...) indicates that this wave does not depend on the random potential at any site.  The simplest distorted wave is,

$$\Phi_{(1, 0, 0, ...)} = \sum_{\alpha} p_1(\varepsilon_{\alpha}) \, \phi_{\alpha}. \tag{3}$$

where the first component of the index is one to indicate that the coefficient of $\phi_{\alpha}$ depends linearly on $\varepsilon_{\alpha}$.  The polynomial $p_1(\varepsilon)$ is of degree one in $\varepsilon$, and constructed to be orthonormal to the polynomial of degree zero, $p_0(\varepsilon)$=1, with respect to integration over $\rho(\varepsilon)$.  The inner product between distorted waves is the average over products of components, which makes these two waves orthonormal provided there are enough sites (infinite for continuous distributions) in the lattice so that the average is equal to the integral of $p_1(\varepsilon)$ over $\rho(\varepsilon)$,

$$\Phi_{(0,0,0,...)} \cdot \Phi_{(1,0,0,...)} = <p_1(\varepsilon_{\alpha})> = 0. \tag{4}$$

The general distorted wave has components whose dependence on the random potential at various sites is given by polynomials $p_n(\varepsilon)$ which are of degree n and are orthonormal with respect to integration over the distribution of site-energies $\rho(\varepsilon)$,

$$\int p_n(\epsilon)\, p_m(\epsilon)\, \rho(\epsilon)\, d\epsilon = \delta_{n,m}. \tag{5}$$

In terms of these polynomials, the general distorted wave is,

$$\Phi_s = \sum_\alpha \{\prod_\beta p_{s(\beta)}(\epsilon_{\alpha+\beta})\}\, \phi_\alpha, \tag{6}$$

where different components of $s$ are indexed by $\beta$, and $\epsilon_{\alpha+\beta}$ is the energy of the orbital at $R_\alpha + R_\beta$. The inner product between two distorted waves is the average over $\alpha$ of the product over $\beta$ of the pairs of polynomials (one from each wave) in the energies of the sites $\epsilon_{\alpha+\beta}$,

$$\Phi_s \cdot \Phi_{s'} = <\prod_\beta p_{s(\beta)}(\epsilon_{\alpha+\beta})\, p_{s'(\beta)}(\epsilon_{\alpha+\beta})>. \tag{7}$$

Provided there are enough sites (usually an infinite number) so that as $\alpha$ varies the distribution of site energies $\{\epsilon_{\alpha+\beta}\}$ is identical to the distribtuion of site-energies $\{\epsilon_\beta\}$ obtained by choosing each site-energy independently from the distribution $\rho(\epsilon)$,

$$\Phi_s \cdot \Phi_{s'} = \prod_\beta <p_{s(\beta)}(\epsilon_{\alpha+\beta})\, p_{s'(\beta)}(\epsilon_{\alpha+\beta})> = \prod_\beta \delta_{s(\beta),s'(\beta)}. \tag{8}$$

In this basis of distorted waves, the matrix-elements of the Anderson Hamiltonian are easy to calculate and take definite values rather than distributions of values, because the averaging in the inner product is equivalent to integration over the products of distributions. The hopping term T is the second term in Eq. 1, and couples neighboring orbitals. When T acts on one of the distorted waves $\Phi_s$ it replaces each $\phi_\alpha$ in Eq. 6 with h times the sum of its neighboring orbitals. Writing the result as a sum of distorted waves $\Phi_{s'}$: for each nearest neighbor translation $\delta$ of the lattice, the component $s'(\beta)$ takes the value of the component $s(\beta+\delta)$,

$$T\, \Phi_s = h \sum_\delta \Phi_{s'}, \text{ where } s'(\beta+\delta) = s(\beta). \tag{9}$$

Note that when **s** is **0**, **s'** and **s** are the same and so $\Phi_0$ is an eigenstate of T with energy $2d$h, as it should be.

The disordered potential V is the first term in Eq. 1 and multiplies an orbital by its energy. The property of orthogonal polynomials which makes them useful for this basis is that multiplying one of them by its argument produces a combination of at most three polynomials, [15],

$$\varepsilon \, p_n(\varepsilon) = b_{n+1} \, p_{n+1}(\varepsilon) + a_n \, p_n(\varepsilon) + b_n \, p_{n-1}(\varepsilon), \tag{10}$$

where the coefficients $\{a_n\}$ and $\{b_n\}$ are determined by the details of $\rho(\varepsilon)$ [15]. So, the action of V on a distorted wave gives,

$$V \, \Phi_{\mathbf{s}} = b_{n+1} \, \Phi_{\mathbf{s+1}} + a_n \, \Phi_{\mathbf{s}} + b_n \, \Phi_{\mathbf{s-1}}, \tag{11}$$

where **1** is the vector index (1,0,0,...) and $\Phi_{\mathbf{s}}$ is taken to be zero if any component of **s** is negative.

This representation of the Anderson Hamiltonian replaces the random energies of orbitals with the vector index **s** which may be interpreted as an internal degree of freeedom for each site, in other words a spin on each site. The number of spin states for each orbital is equal to the number of different values the random potential can take, infinite for the top hat distribution, but two for a binary alloy. Indeed, the relation between the spin Hamiltonian and the random orbital energy is simply that the distribution of orbital energies is the projected density of states for $\Phi_0$ when the hopping term T is zero. Since the disordered potential V only changes the first component of **s**, spins on different sites only interact through T which shifts the spins from one site to another. This transformation of the Anderson Hamiltonian to a basis of distorted waves replaces disorder with interactions, a kind of reverse of the Hubbard-Stratanovich transformation.

A more familiar interpretation of the above transformed model is that T generates hopping of an electron rather than the spins. The random potential V then generates an interaction between the electron and the spins in the sense that the spin on a site can only change when the electron is on the same site. In this picture, $\Phi_0$ is the vacuum state, and $\Phi_1$ is the state with the electron and a single spin excitation at the origin. In subsequent applications of H to $\Phi_1$, T generates electron hopping from one site to another and V generates interactions which can change the spin on the site occupied by the electron. In this interpretation, the transformed model is closely related to the representation of the Anderson model in augmented space [16,17], developed to calculate averaged quantities in disordered systems and derived by explicit averaging. Here, a similar representation is derived for a single realization of the infinite disordered system with the connection between the two approaches being that the single infinite realization of the system includes all infinite realizations as subsystems, and so the distorted waves self-average.

Extended wave functions, whether basis functions or states of the system, have very different properties from localized states or orbitals. In the limit of an infinite system, the number of sites goes to infinity, but remains countable while in this limit the extended states form continua which are uncountable, for example the continuum of Bloch states which comprise a band for a crystal. For the Anderson model, the distorted waves $\{\Phi_s\}$ also form a continuum which can be demonstrated by the one-to-one mapping of distorted waves to the continuum of real numbers in which the components of the vector index **s** are interpreted as digits in a representation whose base is the number of different values the random potential can have at each site.

## 3. Bands in the Transformed Model

Although the random potential prevents the Anderson Hamiltonian from having any geometric symmetries, its matrix-elements between distorted waves display order because they are averages over infinite numbers of site-energies. For example, the same matrix-element h translates the excitation in a wave such as $\Phi_{(0,0,...,0,1,0,...)}$ by a nearest-neighbor lattice vector, and the same is true of waves with more complicated excitations in which case the excitations are all translated by the same nearest-neighbor lattice vector. Similarly, the same matrix-elements $\{a_n\}$ or $\{b_n\}$ couple every wave $\Phi_s$ to itself and to the waves whose first component of $\mathbf{s}$ is greater or smaller by one.

The order in this representation of the Anderson Hamiltonian allows four different kinds of states to be distinguished from one another. The simplest states are localized in both excitation number and components of the index $\mathbf{s}$ of $\Phi_s$; or in other words these states are very like a single distorted wave in that they contain significant contributions from just a few waves. Next there are states which are delocalized either in excitation number, or in index-space. The former are characterized by having their weight distributed over an infinite range of excitation numbers, but in just a few components of $\mathbf{s}$. The latter states contain just a few excitation numbers, but these are spread over an infinite number of components of $\mathbf{s}$. Finally there are states which are distributed over an infinite range of excitation numbers and components of $\mathbf{s}$.

The different kinds of states form bands whose edges, the energies at which states change qualitatively, are the boundaries in the phase diagram for the model. States which contain just a few distorted waves are the extended states of the Anderson model because the distorted waves themselves are extended, while those which are spread over all distorted waves are the localized states of the model because the different distorted waves cancel almost everywhere. At this time it is not clear how the states of mixed character, localized in either excitation number or component of $\mathbf{s}$ and delocalized in the other, should be interpreted in the Anderson model. This may have something to do with the weak transitions discussed in Sec. 4.

In the site-representation, the Anderson Hamiltonian acts on electronic orbitals distributed over a lattice embedded in one, two, or three spatial dimensions; however, the basis of distorted waves is indexed by $\mathbf{s}$ which has an infinite number of components, not just one, two, or three. If the distorted waves are to be identified with sites on a lattice, then the lattice must be embedded in an infinite dimensional space to accomodate the infinite number of components. Althought the dimension of the embedding space goes to infinity, each wave is only coupled to at most 2$d$ other waves by the hopping part of the Hamiltonian, and to at most 2 other waves by the random potential.  The infinite dimensionality of the embedding space allows states of kinds not seen on finite dimensional lattices, but the finite coordination of each site on the infinite dimensional lattice makes it possible to solve the model.

3.1  The Cayley Tree

The simplest example of an infinite dimensional lattice with finite coordination is a Cayley tree[2], used here to illustrate the Anderson Hamiltonian represented in the distorted waves.  Each site on a Cayley tree of coordination Z is coupled to Z neighbors, and the property which makes it a tree is that between any pair of distinct sites there is a unique self-avoiding path from site to coupled site.  Taking one site as the origin, this property can be used to index each site in the Cayley tree by the self-avoiding path to it from the origin.  The origin is the only site reached by a path of zero length and so it is the one site in the zero-th generation; its Z neighbors are each reached by a different path of length one, making up the first generation; each of those has Z-1 neighbors other than the origin, making a total of Z(Z-1) sites in the second generation, each reached by a self-avoiding path of length two; and so on with the Z(Z-1)$^{n-1}$ self-avoiding paths of length n each reaching a different site in the n-th generation.

The Cayley tree has two kinds of states with qualitatively different distributions of weight over the sites of the Cayley tree.  For example, a state at one edge of what might be called the band of localized states has component 1 at the origin, components 1/Z$^{1/2}$ on each site in the first generation, and so on with components 1/[Z(Z-1)$^{n-1}$]$^{1/2}$ on each site in the n-th generation.  The energy of this state is 2(Z-1)$^{1/2}$ in units of the coupling between neighboring sites, and it is localized in the sense that its weight on a each site in a generation decreases exponentially with the number of the generation.  The state at one edge of what might be called the band of extended states has component 1 on every site of the tree.  Its energy is Z in units of the coupling and it is extended in the sense that its weight is distributed equally over every site of the tree.  The localized and extended states of the tree each belongs to a band of similar states which can be constructed by varying the phase relationships between components on neighboring sites. The localized states form a band from $-2(Z-1)^{1/2}$ to $2(Z-1)^{1/2}$, and the extended states form a band from $-Z$ to $Z$, both in units of the inter-site coupling. Note that between $-2(Z-1)^{1/2}$ and $2(Z-1)^{1/2}$ the extended states and localized states are degenerate, not because of any symmetry, but because of their qualitatively different normalizations.

## 3.2 Localized States

Like sites on a Cayley tree, the distorted waves have an infinite number of indices, and so if coupled waves are to be represented by neighboring sites, they must be in an infinite dimensional space. While the expansion of the Anderson Hamiltonian in the distorted waves is not a tree, a wave $\Phi_s$ can be assigned to the n-th generation if n is the minimum power of the Hamiltonian which has a non-zero matrix-element between $\Phi_0$ and $\Phi_s$, which is to say that n is the least number of non-zero matrix-elements of T or V between $\Phi_0$ and $\Phi_s$. Like the Cayley tree, the number of basis elements in the n-th generation increases exponentially with n, but unlike a tree, there are many different self-avoiding paths between any two basis elements.

Like the Cayley tree, the Hamiltonian in distorted waves has states for which the weight on each wave is the same. To find out how these states are distributed on the site-orbitals of the original Anderson model, the distorted waves must be expressed in terms of site-orbitals and summed. Just as plane waves in k-space transform to delta-distributions in direct-space, these states which are delocalized in distorted waves transform to highly localized combinations of site-orbitals because the superpositions of waves cancel at all but exceptional sites.

The first step in calculating the energies of the extremal states of this band is that of determining the asymptotic properties of V, the random potential, in the distorted waves . For an Anderson model with a semi-circular distribution of site energies having width W and centered at zero, the non-zero matrix-elements of V between distorted waves are all W/4, as can be seen from the recurrence relation for Chebychev polynomials of the second kind[15] which are orthogonal with respect to this distribution. The usual distribution for site energies in the Anderson model is the top hat distribution of width W centered at zero, for which Legendre polynomials [15] are the orthogonal polynomials. The recurrence relation for these polynomials gives non-zero matrix-elements of V which are not all W/4, but go to W/4 as the degree of the polynomials goes to infinity. Indeed, for any smooth distribution of site energies with width W, it is shown in Ref. 15 that the polynomials orthogonal with respect to that distribution have a recurrence relation with coefficients approaching W/4 as the degree of the polynomial increases.

The second step in this calculation is to observe that the weight of these states, delocalized in the distorted waves, is dominated by the asymptotic region of V in which the non-zero matrix-elements of V are all W/4 and the hopping matrix-elements are all h. In this part of the matrix, the extremal states of this delocalized band are simple and just like those for the Cayley tree: one extremal state has component 1 on every distorted wave, and the other alternates +1 and -1 on neighboring distorted waves, which it can do because the site lattice is bipartite and so is the representation of V in the asymptotic region. For a lattice of dimension $d$, each distorted wave is coupled to $2d$ neighbors with matrix-elements h and to two neighbors with matrix-elements W/4 in the asymptotic region. The constant state has energy $2d$h+W/2 where $2d$ is the coordination of the site lattice, and the alternating state has energy -$2d$h-W/2. These are just Lifshitz edges for the localized states of the Anderson model, consistent with the interpretation of the delocalized states in the basis of distorted waves as the localized states in the site-orbitals.

3.3 Extended States

Like the Cayley tree, the Anderson Hamiltonian in the distorted waves also has a band of states which contain just a few waves, localized in the wave-basis; but unlike the Cayley tree, the extremal energies of this band cannot be calculated exactly. States which are localized in the waves are extended in the original basis of site-orbitals simply because the waves themselves are extended states, like the constant state but distorted by the random potential.

The simplest method for estimating the extremal energies of this band of states which are localized in the waves but extended in the orbitals, is to tridiagonalize the Anderson Hamiltonian. While this tridiagonalization can be done numerically, the simple variational approximation developed below makes it easy to understand the nature of these states. This approximate tridiagonal basis is $\{u_n\}$ where $u_n$ has equal coefficients on each wave in the n-th generation of waves as defined in Sec. 3.2. The first state $u_0$ is just $\Phi_0$ the constant wave. The second state $u_1$ is $\Phi_{(1,0,0,...)}$ and the third state $u_2$ is the symmetric combination of all the $\Phi_s$ which are coupled to $\Phi_{(1,0,0,...)}$ by T or V, and so on so that $u_n$ is the symmetric combination of all the $\Phi_s$ for which n is the smallest power of the Hamiltonian with a non-zero matrix-element between $\Phi_s$ and $\Phi_0$. The projection of the Hamiltonian on this basis $\{u_n\}$ is tridiagonal, but only an approximate tridiagonalization because the $\{u_n\}$ are not complete in the smallest invariant subspace of states containing $\Phi_0$. However because it is a projection, this approximation is variational and only reduces the widths of bands. More simply, the approximate mobility edges of the approximation are upper (lower) bounds for the lower (upper) exact mobility edges.

The approximation made in this projection is that each of the $\{u_n\}$ is supported on a single generation of the extended basis and is a symmetric combination of basis elements within that generation. Numerical tridiagonalization of the Anderson Hamiltonian [18] produces a basis like the $\{u_n\}$, but not localized on a single generation, and not symmetric within each generation. However, at band edges where the weights of states do not vary much from one element of the extended basis to the next, and for W comparable to h, this is a good approximation.

The projection of the Anderson Hamiltonian from the basis of distorted waves to the $\{u_n\}$, which are localized on single generations, produces a tridiagonal matrix because the Hamiltonian only couples elements in neighboring generations. For an Anderson model with sites on a *d*-dimensional analog of the cubic lattice, a typical distorted wave in the asymptotic region is coupled by W/4 to a wave in the preceding generation and to one in the succeeding generation where the first component of **s** changes by -1 and +1 respectively. This typical wave is coupled by the hopping matrix-element h to *d* waves in the previous generation and *d* in the succeeding generation accounting for the 2*d* nearest-neighbor translations of **s**. The simplest calculation of the tridiagonal matrix-elements is to multiply $u_n$ by the Hamiltonian and normalize the projections of the result on $u_{n-1}$ and $u_{n+1}$ which gives zero for the diagonal elements of the tridiagonal matrix and $[(W/4)^2 + d^2 h^2]^{1/2}$ for the off-diagonal elements in the asymptotic region.

The conclusion of this calculation is that in the basis of distorted waves the Anderson Hamiltonian has a band of localized states between $\pm[(W/2)^2 + 4d^2 h^2]^{1/2}$, corresponding to a band of states extended in the site-orbitals. How this result is related to the results of tridiagonalizations in the basis of site-orbitals and to the energies of mobility edges is discussed below.

4. Mobility Edges

In two dimensions, previous analytic and numerical tridiagonalizations have found sharp localization edges [7,8,9,18] in disagreement with other methods; while in three dimensions for small disorder, both analytic[7] and numerical[18] tridiagonalizations agree with other methods in finding sharp mobility edges. In three dimensions with small disorder the band of extended (in the site-orbitals) states which arises in many different approaches is clearly the same as the band found in the above variational calculation, and the quantitative agreement between these results is discussed below. However, there are several issues which arise from this. The first is that the width of this band of extended states continues to increase in three-dimensional systems at disorders far beyond the Anderson transition, in conflict with other approaches [3,4,5,6] and in seeming contradiction to the existence of the Anderson transition. The second is that this band of extended states also occurs in two-dimensional systems where many methods find only exponentially localized states, and where local tridiagonalizations (starting with single site-orbitals) show power-law localized states, again a seeming paradox. The third issue is that this band of extended states also occurs in one-dimensional system where neither tridiagonalizations starting from a single site-orbital nor any other method has shown any sign of mobility edges. The remainder of this Sec. discusses these issues.

4.1 Comparison with Previous Tridiagonalizations

Recently Arnold [18] used a parallel numerical implementation of the recursion method to tridiagonalize Hamiltonians for Anderson models with up to $4 \times 10^9$ sites on square and cubic lattices, samples more than 1000 times the size used in previous numerical studies [5,23] using the Lanczos method. In contrast to the work described above, these calculations begin the tridiagonalization with a single site-orbital rather than a constant wave. As with previous work[8,9], the tridiagonal matrix-elements obtained from these samples have transmittances for escape from the initial orbital with consistent power-law or exponential behavior over the last several hundred elements which we take to be their asymptotic behavior. The mobility edges appear as strong features in plots of the energy-dependence of this transmittance. The energies of these mobility edges are shown in Figs. 1 and 2 together with error estimates for their energies for various disordered square and cubic lattices.

The above calculations of extremal energies for the band of states which are localized in the distorted waves and so extended in the site-orbitals, give variational bounds on the mobility edges for square and cubic lattices, respectively $\pm[(W/2)^2+16h^2]^{1/2}$ and $\pm[(W/2)^2+36h^2]^{1/2}$, shown as dashed lines in Figs. 1 and 2. In most cases the analytic approximation lies on the interior extreme of the numerical estimates, and in a few cases, just inside the extreme numerical estimate. This is consistent both with the above analytic results being variational bounds on the trajectories of the mobility edges, and with the variational tridiagonalization being a good approximation.

In earlier analytic work Haydock [7] found equations for the trajectories of mobility edges in two and three-dimensional Anderson models for small disorder. The models in that work differ from those used here in that the electronic states at zero disorder are plane-waves rather than tight-binding states, and so the two models have different densities of states. In order to compare the results, the models can be related by the widths of the bands for zero disorder and the width of the distribution of potentials. For the three-dimensional model [7] the leading power of W in the mobility edge trajectory is $W^2/36h$, compared to $W^2/48h$ here; which is probably due to differences in the effective masses reflecting the different shapes of the densities of states. In the two-dimensional model [7] the mobility trajectory has an exponential cusp instead of a square-root cusp for zero disorder  This is a qualitative difference between the two results and is probably due to the approximate tridiagonalization's failure to reproduce the logarithmic singularities in the densitiy of states at zero disorder. This conclusion is supported by other numerical work [8] on the two-dimensional Anderson model which agrees much better with the previous analytical work at low disorder [7] than with the current work.

Taking into account the differences between the various models, methods, and approximations used, the current work and previous tridiagonalizations are consistent with one another.

## 4.2  Interpretation of Tridiagonalizations

Qualitative distinctions such as between localized and extended states are properties of infinite systems, or more precisely, the limits of finite systems as they become large.  While many different kinds of states are possible in infinite systems, the two kinds which are important in the Anderson model are localized states, which have significant weight on only a finite number of sites, and extended states, whose weight is distributed over an infinite number of sites.

Tridiagonalizations such as in Sec. 3 and in Refs. 7, 8, and 18 project the Hamiltonian on the states having greatest relative overlap with the starting state.  If there are two qualitatively different states at the same energy and both overlap the starting state, then the state which contains more of the starting state is spanned by the tridiagonal basis which is orthogonal to the one with the smaller relative overlap.  This property of tridiagonalization is discussed more thoroughly in Ref. 19.  For the tridiagonalization in Sec. 3 the starting state is a single distorted wave, so for energies within the band of wave-like states, the tridiagonalization projects onto these wave-like states because they have the greatest relative overlap with the starting state. Even if there are also orbital-like states at the same energies as the wave-like states, their relative overlap is zero and so the tridiagonal basis is orthogonal to them.  Outside the band of wave-like states, the states with the greatest overlap are the orbital-like states, so they occur as divergent sums of the $\{u_n\}$. In Refs. 7, 8, and 18 the starting states are site-orbitals, so by the same reasoning as above, the tridiagonalization projects onto localized states and only includes extended states when there are no ordinary localized states at the same energies.

For large disorder the presence of bands of wave-like states in tridiagonalizations beginning with a single distorted wave, and bands of orbital-like states in tridiagonalizations beginning with a single site-orbital, indicate that there is degeneracy between extended and localized states in these systems. This is surprising because a disordered potential has no symmetries and one might think that lack of symmetry was enough to prevent such degeneracies. However, this is not necessarily so as can be seen by considering the total weight of a site-orbital in the band of extended states. If this weight goes to zero sufficiently strongly with system size, then the two kinds of states do not couple, despite the lack of symmetry, and they can be degenerate because they differ qualitatively.

The agreement in Figs. 1 and 2 between numerical estimates of the energies of mobility edges and the analytical approximation derived in the previous Sec. supports the interpretation that the mobility edges are the edges of the band of wave-like states constructed from the distorted waves. This is consistent with previous calculations for two dimensions using tridiagonalization, and with most calculations for three dimensions using various methods. However, the persistence of the mobility edges, and the band of states which are localized in the distorted waves, to large disorder raises the question of how these extended states affect the propagation of an electron which starts on a single site.

While it is clear that for large disorders an electron starting on a single site cannot escape from some finite neighborhood of that site, Ref. 18 shows strong evidence of singularities in the energy-dependence of the propagation of an electron starting from a single site in two and three dimensions; and these singularities agree with the edges of the band of extended states obtained in Sec. 3.3. The explanation of this seeming paradox is that in two and three dimensions, tridiagonalizations starting with a single site produce basis elements which are less localized with increasing index until in the infinite limit the tridiagonal basis elements are extended states. Since these tridiagonal bases approach extended states, they couple to the band of extended states, and the singularities at the edges of these extended bands induce singularities in the localized states with the same energy - in other words, there is a kind of coupling between the two kinds of states which is too weak to break the degeneracy. For large disorders, the mobility edges separate energies where electrons are exponentially localized on single sites from those where they are only power-law localized, or for even greater disorders the mobility edges occur where the dependence of localization length on energy is singular.

The situation in one dimension is different in that no method produces mobility edges in the propagation of electrons starting from a single site, while the band of extended states can still be constructed as above. Unlike two and three dimensions, in one dimension the tridiagonal basis starting with a single site-orbital never becomes extended, even for an ordered system. Since all elements of the tridiagonal basis are localized on just a few sites, these tridiagonalizations never detect the band of extended states. In this sense, the localized and extended states are totally decoupled for finite disorder in one dimension while in two and three dimensions the disorder must become infinite to achieve total decoupling.

5. Anderson Transitions

It is argued above that the mobility edges continue to separate with increasing disorder, contrary to the picture in which the mobility edges meet at the center of the band at the critical disorder for the Anderson transition. If the mobility edges never meet, then the Anderson transition must occur by a different mechanism, and it is argued below that the mechanism is decoupling of the entire band of extended states from the site-orbitals, which can take place in several steps producing multiple Anderson transitions.

5.1 Decoupling

　　The simplest approach to the relation between mobility edges and Anderson transitions is through the asymptotic properties of the local tridiagonalizations, those starting with a single site-orbital, in contrast to the extended tridiagonalizations, those starting with a constant wave, discussed in Sec. 4.2. A local tridiagonalization transforms the $d$-dimensional random lattice into a one-dimensional, semi-infinite chain of renormalized sites with only nearest neighbor interactions. Simple dimension counting requires that the individual elements of the tridiagonal basis be concentrated in ($d$-1)-dimensional regions of the random lattice, and this is exactly what is seen in numerical tridiagonalizations [18]. In the asymptotic part of the tridiagonalization the number of sites in the ($d$-1)-dimensional regions approaches infinity for $d$ greater than one, and this determines the asymptotic properties of the tridiagonal representation of the Hamiltonian, which in turn determines the asymptotic properties of the states.

　　For a local tridiagonalization, the most prominent asymptotic features of the tridiagonal matrix are the limit points of its elements. For $d$ greater than one, the number of independent site-energies contributing to each matrix-element approaches infinity, so the distribution of values for each matrix-element approaches a delta-distribution according to the central limit theorem. The limit points of the matrix-elements are then the values where the delta-distributions are non-zero. For the Anderson model in Eq. 1 with a symmetric distribution of site-energies on a bipartite lattice of dimensionality greater than one, the limit point of the diagonal elements of the tridiagonal matrix is zero [20]. As the tridiagonal basis elements approach extended states, they must span the band of extended states found in the previous Sec., and so the limit point of the off-diagonal elements is approximately $[(W/4)^2 + d^2 h^2]^{1/2}$. Provided that the tridiagonal basis elements approach extended states, these limit points correspond to the mobility edges and because they are the only limit points, there are no other energies where the properties of extended states can become singular.

　　Next to the limit points of the tridiagonal matrix-elements, the most prominent feature of their asymptotic properties are the fluctuations of matrix-elements around these limit points, and it is these which govern the Anderson transitions. For energies within the band of extended states, the random component of these fluctuations scatters electron waves back toward the initial site, and although other random fluctuations can scatter parts of these waves forward again, multiple scattering from different sequences of sites adds incoherently and can be neglected. In ordered systems the structure of the lattice produces coherent fluctuations in the tridiagonal matrix-elements which persist in disordered systems and do not contribute to decoupling of the localized and extended states.

The coupling between a single site and the extended states is given by the transmittance of the tridiagonal matrix for an electron wave starting at the zero-th element of the tridiagonal basis, which is a single site-orbital, and going to infinity in the tridiagonal basis, which is an extended state for $d$ greater than one and finite disorder. Within the band of extended states, the random fluctuation of each matrix-element reduces the logarithm of the mean transmittance by an amount proportional to the square of the fluctuation, see further Ref. 7. If the sum of these reductions in transmittance diverges, then the extended states are decoupled from the sites and no electrons starting on a single site can escape to infinity. If this sum converges then the extended states are coupled to single sites and electrons starting on single sites can escape to infinity.

## 5.2 Anderson Transitions at Zero Disorder

The simplest examples of Anderson transitions are those which occur at zero disorder. For one-dimensional systems, however small the disorder, the tridiagonal basis elements are zero-dimensional, meaning that the number of independent site-energies contributing to each matrix-element does not diverge in the asymptotic part of the matrix. As a result, the fluctuations in the tridiagonal matrix-elements do not go to zero, so the band of extended states is completely decoupled from the asymptotic part of the tridiagonal matrix, and there are no limit points around which the matrix-elements fluctuate. However the states coupled to individual sites can be obtained by diagonalization of the tridiagonal matrix [18] and, as found by others [2] are exponentially localized with a localization length which varies smoothly with energy. This Anderson transition occurs throughout the band at zero disorder and is between extended states for no disorder and exponential localization near individual sites for any non-zero disorder.

Another example of an Anderson transition at zero disorder occurs in two dimensions. In this case the tridiagonal basis elements are spread over one-dimensional shells, and so of order n independent site-energies contribute to the n-th diagonal or off-diagonal element of the tridiagonal matrix. From the central limit theorem, the mean square fluctuation is proportional to $1/n$ and so the tridiagonal matrix-elements have limit points corresponding to a band of extended states at infinity, and for energies within that band the back scattering sum diverges logarithmically [7] leading to a transmittance which decreases as a negative power of the distance from the initial site, power-law localization in the vicinity of the initial site for energies within the band of extended state. Outside the band of extended states, states decay exponentially with distance from a single site, with a localization length which varies as the root of the energy relative to the mobility edge just as they do outside any other band edge, and the edges of the band of extended states are the mobility edges which in this case separate exponential localization from power-law localization. The Anderson transition takes place across the entire band of extended states at zero disorder because the divergence of the back scattering series for the transmittance is independent of energy within the band. The power-law with which electrons are localized near single sites does vary with energy having the weakest localization at the center of the band and the strongest localization at the edges of the band where the power-law diverges.

In three dimensions the n-th element of the tridiagonal basis is spread over a two-dimensional shell which includes of order $n^2$ independent sites. From the central limit theorem, the mean square fluctuation in the matrix-elements decreases as $1/n^2$ for small disorders, so for energies within the band of extended states the back scattering sum converges and individual sites are coupled to the extended states at infinity. For the same reasons that states outside any other band edge are localized, states at energies outside the mobility edges decrease exponentially with distance from some site with a localization length which varies as the root of the energy relative to the mobility edge.  Although electrons can escape to infinity at any energy within the band of extended states, the rate at which they do so varies with energy and goes to zero at the mobility edges.

5.3  Anderson Transitions at Non-zero Disorder

There must be additional Anderson transitions at non-zero disorders separating exponential localization at infinite disorder from the power-law localization found for weak disorder in two dimensions and the delocalization found for weak disorder in three dimensions. As disorder increases, the tridiagonal basis elements become concentrated on the site-orbitals with the most extreme energies, increasing the fluctuations in tridiagonal matrix-elements because there is less averaging.  In the limit of infinite disorder the tridiagonal basis elements are concentrated on single sites [13] and the matrix-elements cease to have single limit points, like the one-dimensional case.

While there are no analytic approximations currently available to describe the fluctuations in tridiagonal matrix-elements at intermediate disorders, the analysis used above applies to the limit points of these matrix-elements.  The dependence on n of fluctuations in the n-th tridiagonal matrix-element determines whether the back scattering series converges or not, and because the tridiagonal matrix-elements are independent of energy, whether or not the series converges is also independent of energy.  Consequently Anderson transitions occur across the entire band at the same critical disorder for which the convergence properties of the back scattering series change.

In three dimensions the disorder at which this series ceases to converge and becomes logarithmically divergent corresponds to an Anderson transition from delocalization to power-law localization in the vicinity of a single site.  While the kind of localization changes throughout the extended band at the same disorder, the strength of localization which is the power-law itself varies across the band, diverging at the mobility edges.  Evidence of this Anderson transition can be seen in the numerical work of Arnold [18] at a disorder W of about 14h±2h.

Arnold's results [18] differ from previous work using different numerical methods [21,22,23] in that the critical disorder is lower, and that the transition is from extended to power-law localized states rather than from extended to exponentially localized states. As can be seen in Fig. 2, the combination of the mobility edges at the band extrema and the transition line connecting the mobility edges at W=14 is not so different from the picture of mobility edges curving around to meet at the band center. For disorders greater than the critical disorder, Arnold's work indicates a singular region in which the power-law varies with disorder, much like the variation of the power-laws for correlations in the Kosterlitz-Thouless transition [24]. There is no evidence in Arnold's work of exponential localization at the band center up to disorders as high as W=24.

In two dimensions Arnold [18] finds both mobility edges and an Anderson transition separating power-law localized states from exponentially localized states, while other numerical methods do not. Figure 1 shows the Anderson transition at 16h±2h connecting the Lifshitz edges rather than the mobility edges because this transition is a singular change in the exponential localization length rather than in a power law. For disorders less than the critical disorder, there is again a singular region in which the power-laws vary as they do in the Kosterlitz-Thouless transition.

## 5.4 Unbounded Disorder

Since field theoretic methods have been used widely to study localization, and for technical reasons the random potentials in such work have Gaussian distributions, it is worth applying the methods used above for bounded distributions to unbounded distributions such as Gaussians. Although Gaussians seem little different from bounded distributions, the Anderson models for Gaussians and other unbounded distributions is unlike a bounded distribution in that they each have an infinite number of sites with energies greater than any given finite energy. It turns out, as is explained below, that the consequences of unbounded distributions are very serious and these consequences should be apparent in the field theoretic models.

Hermite polynomials [15] are orthogonal with respect to integration over Gaussian distributions, and their recurrence relation[15] gives matrix-elements for V in the extended basis. Because of the unbounded property of Gaussian disorder, these matrix-elements grow as the root of the index of the polynomial rather than approaching a finite limit as for the top hat distribution. The consequence of this is that the matrix-elements from the variational tridiagonalization of the Anderson Hamiltonian for this distribution also grow as the root of the index rather than going to a constant. Since it is the asymptotic part of the tridiagonalization which determines the singular energies, the mobility edges for Gaussian disorder are at infinite energy just as they would be for a top hat distribution of infinite width. Once again there are no other limits of the tridiagonal matrix-elements corresponding to singularities at some finite energy, so the band of extended states is infinite in width. Since other unbounded distributions such as the Cauchy distribution are wider than the Gaussian, they also lead to mobility edges at infinity.

Although the mobility edges are at infinite energy for unbounded disorder, it is the fluctuations in the matrix-elements of a local tridiagonalization which determine the critical width at which the Anderson transition occurs for these models. For one or two dimensions and a Gaussian distribution of site-energies, the Anderson transition must occur at zero disorder for the same reasons it occurs for other distributions at zero disorder. In three dimensions the question is more delicate.

For unbounded disorder in any infinite system there are an infinite number of sites whose orbitals have greater energy than any given ε which is taken here to be much greater than |h|. The disorder can be reduced by approximating the energies of orbital which are less than ε by zero. So the simplified model of unbounded disorder has an infinite number of orbitals, albeit thinly distributed, with energies greater than ε, and the remaining orbitals all having energy zero. Each of the high energy orbitals binds a state with energy greater than ε which must also be present in the localized tridiagonalization of the model. To a good approximation, as is argued in Ref. 13 each extreme orbital becomes a basis element of any localized tridiagonalization of this model, and so the tridiagonal matrix has extreme elements, ε or greater, randomly distributed along its diagonal. The tridiagonal matrix is equivalent to a semi-infinite, one-dimensional system, and so this random distribution of extreme potential fluctuations exponentially decouples the site-orbitals from the infinite band of extended states constructed in the extended basis. As a consequence the Anderson transition for unbounded disorder occurs at zero disorder in all dimensions including three[7].

## 6. Electrons in Disordered Potentials

Representation of the Anderson model in an extended basis reveals several surprises in addition to what has already been learned about this model. While it has been thought that the Anderson transition occurs when mobility edges meet, it now seems that this is not the case. In disordered systems, mobility edges are like the band edges of a crystal in that states outside the edges are exponentially localized because their energies lie outside the band of extended states, while at the Anderson transition scattering off fluctuations in the potential decouples individual site-orbitals from extended states which are degenerate with them. Whereas mobility edges are so like crystalline band edges that they even have effective masses [8], the Anderson transition is something akin to the Kosterlitz-Thouless transition [24] being a continuum of transitions characterized by power laws.

The Anderson model displays a rich variety of behavior because it is infinite in a way crystalline models are not, and yet more readily solved than models of interacting systems. The translation symmetry of a crystal reduces its infinite number of degenerate basis orbitals to a finite number of inequivalent orbitals, but disorder breaks this equivalence restoring the state-space to one of infinite near degeneracy. Once the number of inequivalent nearly degenerate states is infinite, the system can have phases distinguished by the asymptotic behavior of its eigenstates which for disordered systems can vary from extended to exponentially localized through all sorts of intermediate behavior such as power-laws. Of course the state-spaces of interacting systems have infinite numbers of inequivalent configurations, but they are much more difficult to analyze because the number of different configurations grows exponentially with the size of the system rather than linearly as in the Anderson model where the number of orbitals is proportional to the size of the system.

While electrons always interact, single-particle states near the Fermi level in normal metals have arbitrarily long life-times. The Anderson model describes how these long-lived single-particle states can be localized by disorder, and hence how disorder can cause a metal-insulator transition. Of course, for interacting systems localization of metallic electrons by disorder enhances the interactions which leads to further localization. While interactions may drive the transition to lower disorder, it seems that the interactions cannot change the transition qualitatively on the metallic side. However, on the insulating side of the transition, long-ranged interactions make qualitative changes in the transition such as converting power-law localized states into exponentially localized states, or even producing gaps in the electronic excitation spectrum.

Acknowledgments

This work was supported in part by the University of Oregon Foundation, and by the National Science Foundation under DMR- 9319246 from the Materials Theory Program and under STI-9413532 from the Office of Science and Technology Infrastructure.

**Figure Captions**

1.  Numerical results [18] for mobility edges (data points) and the Anderson transition (alternating dashes) compared with analytic trajectories for the mobility edges (short dashes) and Lifshitz edges (long dashes) in the two-dimensional Anderson model.

2.  Numerical results [18] for mobility edges (data points) and the Anderson transition (alternating dashes) compared with analytic trajectories for the mobility edges (short dashes) and Lifshitz edges (long dashes) in the three-dimensional Anderson model.

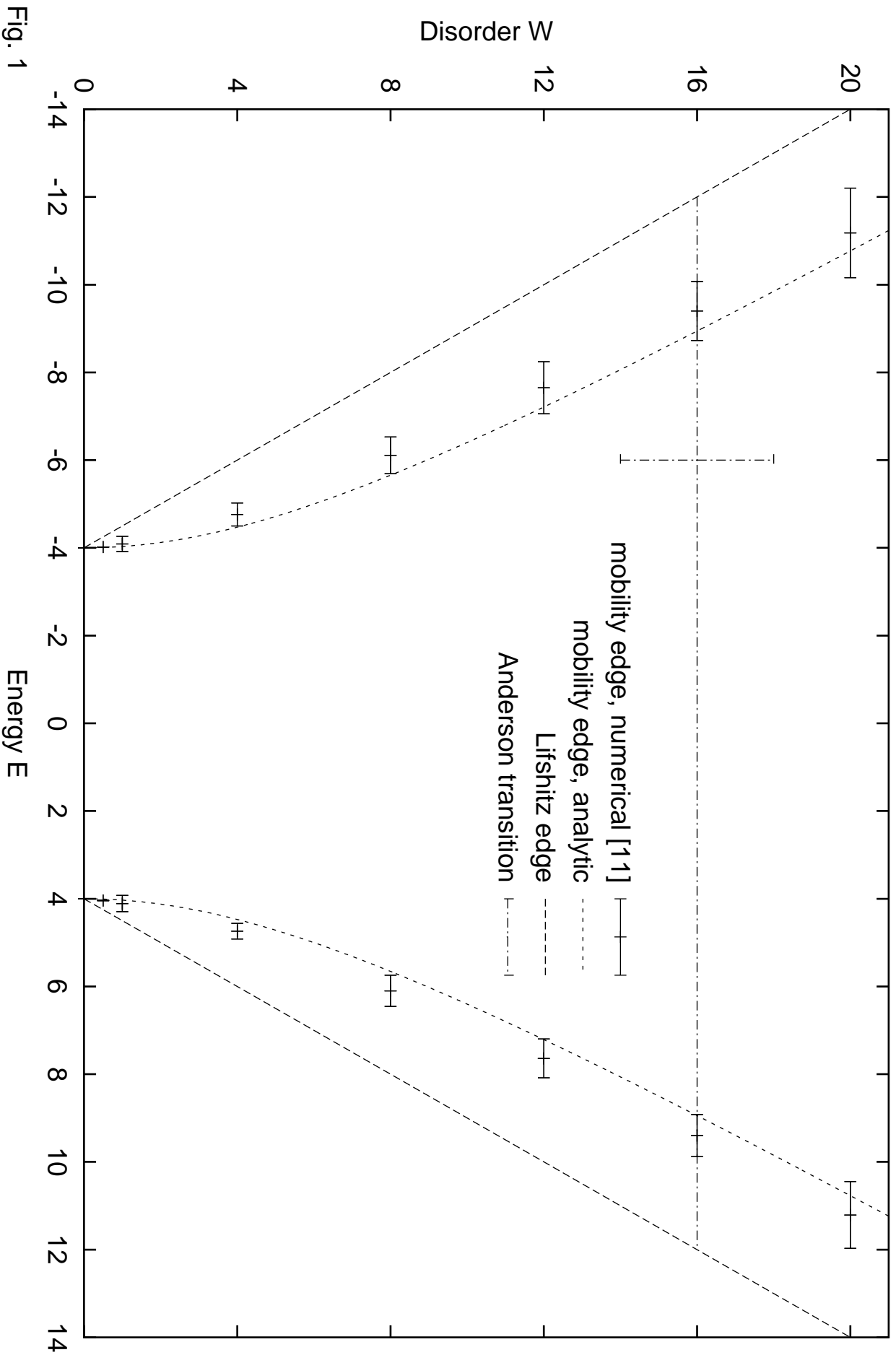

Fig. 1

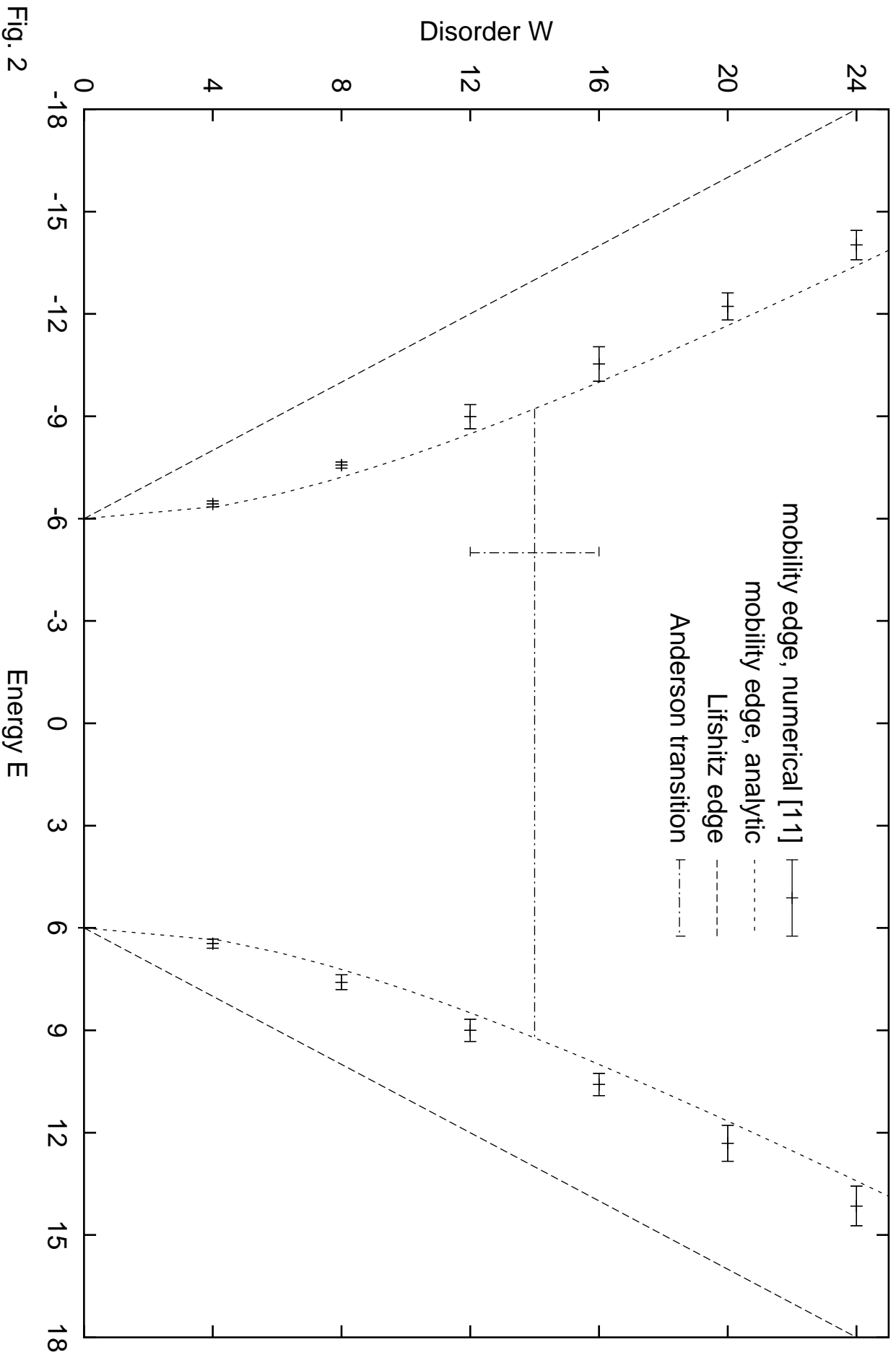

Fig. 2

Mobility edge trajectory in 3D

Disorder W

Energy E

mobility edge, numerical [11]
mobility edge, analytic
Lifshitz edge
Anderson transition